\documentclass[pra,showpacs,showkeys,nofootinbib]{revtex4}

\usepackage{bm}
\usepackage{graphicx}
\usepackage{amssymb}
\usepackage{amsmath}
\usepackage{amsthm}
\usepackage{mathrsfs}
\DeclareSymbolFont{AMSb}{U}{msb}{m}{n}
\DeclareMathSymbol{\N}{\mathbin}{AMSb}{"4E}
\DeclareMathSymbol{\Z}{\mathbin}{AMSb}{"5A}
\DeclareMathSymbol{\R}{\mathbin}{AMSb}{"52}
\DeclareMathSymbol{\Q}{\mathbin}{AMSb}{"51}
\DeclareMathSymbol{\I}{\mathbin}{AMSb}{"49}
\DeclareMathSymbol{\C}{\mathbin}{AMSb}{"43}
\DeclareMathSymbol{\F}{\mathbin}{AMSb}{"46}
\DeclareMathSymbol{\E}{\mathbin}{AMSb}{"45}

\begin{document}

\title{Tight informationally complete quantum measurements}

\author{A. J. Scott}
\email{ascott@qis.ucalgary.ca}
\affiliation{Institute for Quantum Information Science, University of Calgary, Calgary, Alberta T2N 1N4, Canada}

\begin{abstract}
We introduce a class of informationally complete positive-operator-valued measures which are, in analogy with a tight 
frame, ``as close as possible'' to orthonormal bases for the space of quantum states. These measures are distinguished by an 
exceptionally simple state-reconstruction formula which allows ``painless'' quantum state tomography. Complete sets of 
mutually unbiased bases and symmetric informationally complete positive-operator-valued measures are both members of 
this class, the latter being the unique minimal rank-one members. Recast as ensembles of pure quantum states, 
the rank-one members are in fact equivalent to weighted 2-designs in complex projective space. These measures are 
shown to be optimal for quantum cloning and linear quantum state tomography.  
\end{abstract}


\maketitle

\theoremstyle{plain}
\newtheorem{thm}{Theorem}
\newtheorem{lem}[thm]{Lemma}
\newtheorem{cor}[thm]{Corollary}
\newtheorem{prp}[thm]{Proposition}
\newtheorem{con}[thm]{Conjecture}

\theoremstyle{definition}
\newtheorem{dfn}[thm]{Definition}
\newtheorem*{dfn11p}{Definition $\bf{11'}$}

\theoremstyle{remark}
\newtheorem*{rmk}{Remark}
\newtheorem{exm}{Example}

\def\rank{\operatorname{rank}}
\def\tr{\operatorname{tr}}
\def\ket#1{|#1\rangle}
\def\bra#1{\langle#1|}
\def\ketbra#1{| #1 \rangle\langle #1 |}
\def\braket#1#2{\langle #1 | #2 \rangle}
\def\braketb#1#2{\big\langle #1 \big| #2 \big\rangle}
\def\Ket#1{|#1)}
\def\Bra#1{(#1|}
\def\KetBra#1{| #1 )( #1 |}
\def\BraKet#1#2{( #1 | #2 )}
\def\Ketb#1{\big|#1\big)}
\def\Brab#1{\big(#1\big|}
\def\KetBrab#1{\big| #1 \big)\big( #1 \big|}
\def\BraKetb#1#2{\big( #1 \big| #2 \big)}

\def\L{\mathcal{L}}
\def\H{\mathcal{H}}
\def\End{\operatorname{End}}
\def\muu{\mu_{\scriptscriptstyle\mathrm{H}}}
\def\Pisym{\Pi_\mathrm{sym}}
\def\Pisymt{\Pi_\mathrm{sym}^{(t)}}
\def\dsym{d_\mathrm{sym}}
\def\dsymt{d_\mathrm{sym}^{(t)}}
\def\d{\mathrm{d}}
\def\Qd{\operatorname{Q}(\C^d)}
\def\Md{\operatorname{M}(\C^d)}
\def\Pd{\operatorname{P}(\C^d)}
\def\Hd{\operatorname{H}_0(\C^d)}
\def\Htwo{\operatorname{H}_0(\C^2)}
\def\Ih{\mathrm{\bf I}_{\scriptscriptstyle{\mathrm{H}_0}}}
\def\Is{\mathrm{\bf I}}
\def\Tr{\operatorname{Tr}}
\def\Ft{{F'}_{\!\!\!\tau}}
\def\Pt{P}
\def\Rt{R}
\def\FFt{\mathcal{F}}
\def\Ptr{\bm{\Pi}_0}

\section{Introduction}

The retrieval of classical data from quantum systems, a task described by quantum measurement theory, 
is an overlooked -- though important -- component of quantum information processing~\cite{nielsen}.
The ability to precisely determine a quantum state is paramount to tests of quantum information processing 
devices such as quantum teleporters, key distributers, cloners, gates, and indeed, quantum computers. Quality assurance
requires a complete characterization of the device, which is gained through knowledge of the 
output states for a judicious choice of input states. 

The outcome statistics of a quantum measurement are described by a positive-operator-valued measure (POVM) \cite{davies,holevo,kraus,busch2}.
An {\em informationally complete POVM (IC-POVM)\/}~\cite{prugovecki,busch4,busch,hellwig,dariano,dariano2,flammia,weigert} is one with the 
property that every quantum state is uniquely determined by its measurement statistics. A sequence of measurements on copies 
of a system in an unknown state, enabling an estimate of the statistics, will then reveal the state.
This process is called {\em quantum state tomography\/}~\cite{paris}. Besides this practical 
purpose, IC-POVMs with special properties are used for quantum cryptography~\cite{renes2}, quantum fingerprinting~\cite{scott}, 
and are relevant to foundational studies of quantum mechanics~\cite{fuchs,caves2,konig2}. 

This article introduces a special class of IC-POVMs which are, in analogy with a tight frame~\cite{christensen,daubechies2,casazza}, 
``as close as possible'' to orthonormal bases for the space of quantum states. These IC-POVMs will be called {\em tight IC-POVMs\/}. 
They allow ``painless''~\cite{daubechies} quantum state tomography through a particularly simple state-reconstruction formula. The unique minimal rank-one members 
are the symmetric IC-POVMs (SIC-POVMs)~\cite{renes}. Complete sets of mutually unbiased bases (MUBs)~\cite{ivanovic,wootters} 
also form tight IC-POVMs, and in fact, recast as ensembles of pure quantum states, the tight rank-one IC-POVMs are equivalent to weighted 
2-designs in complex projective space. These IC-POVMs are shown to be optimal for linear quantum state tomography and measurement-based 
quantum cloning. 

The article is organized as follows. In the next section we will introduce the notion of a $t$-design in complex projective space. 
Such combinatorial designs have recently aroused interest from the perspective of quantum information theory~\cite{zauner,barnum,renes,klappenecker,hayashi,dankert}.
In Sec.'s~\ref{framesec} and \ref{icpovmsec} we will revise the concepts of operator frames and informational completeness, 
respectively, and then in Sec.~\ref{tightsec}, introduce the 
tight IC-POVMs. We will show in what sense the entire class of tight rank-one IC-POVMs can be considered optimal in 
Sec.'s~\ref{estimatesec} and \ref{optimalsec}, where respectively, linear quantum state tomography and measurement-based cloning 
is investigated. Finally, in Sec.~\ref{concludesec} we summarize our results. Finite dimensional Hilbert spaces are assumed 
throughout the article.

\section{Complex projective designs}
\label{designsec}

The extension of spherical $t$-designs~\cite{delsarte} to projective spaces was first considered by 
Neumaier~\cite{neumaier}, but for the most part studied by Hoggar~\cite{hoggar,hoggar2,hoggar25,hoggar3}, and, 
Bannai and Hoggar~\cite{bannai,bannai2}. For a unified treatment of designs in terms of metric spaces 
consult the work of Levenshtein~\cite{levenshtein05,levenshtein,levenshtein2}. Our interest lies with the 
complex projective space $\C P^{d-1}$ of lines passing through the origin in $\C^d$. In this case each 
$x\in\C P^{d-1}$ may be represented by a unit vector 
$\ket{x}\in\C^d$ (modulo a phase), or more appropriately, by the rank-one projector $\pi(x)\equiv\ketbra{x}$. 
We will use both representations in this article.
Roughly speaking, a complex projective $t$-design is then a finite subset of $\C P^{d-1}$ with the property that 
the discrete average of a polynomial of degree $t$ or less over the design equals the uniform average. Many
equivalent definitions can be made in these terms (see e.g. \cite{neumaier,hoggar,levenshtein05,konig}). 
In the general context of compact metric spaces, for example, Levenshtein~\cite{levenshtein,levenshtein2} calls a finite 
set $\mathscr{D}\subset\C P^{d-1}$ a complex projective $t$-design if 
\begin{equation}
\frac{1}{|\mathscr{D}|^2}\sum_{x,y\in\mathscr{D}}f\!\left(|\braket{x}{y}|^2\right)\;=\;
\iint_{\C P^{d-1}}\d\muu(x)\d\muu(y)\,f\!\left(|\braket{x}{y}|^2\right)
\label{predesign}\end{equation}
for any real polynomial $f$ of degree $t$ or less, where $\muu$ denotes the unique unitarily-invariant 
probability measure on $\C P^{d-1}$ induced by the Haar measure on $\mathrm{U}(d)$. In the current context
we deem it appropriate to make a more explicit definition of a $t$-design which is specialized to 
complex projective spaces. With this in mind, let
$\Pisymt$ denote the projector onto the totally symmetric subspace of $(\C^d)^{\otimes t}$ and consider the 
following simple fact.

\begin{lem}\label{lemsym}
\begin{equation}
\int_{\C P^{d-1}}\d\muu(x)\,\pi(x)^{\otimes t}\;=\;\tbinom{d+t-1}{t}^{-1}\,\Pisymt \label{symavg}\;.
\end{equation}
\end{lem}
\begin{proof}
Use Schur's Lemma. The LHS of Eq.~(\ref{symavg}) is invariant under all unitaries $U^{\otimes t}$ which act irreducibly on the 
totally symmetric subspace of $(\C^d)^{\otimes t}$.  
\end{proof}

By considering the monomial $|\braket{x}{y}|^{2t}=\tr\left[\pi(x)^{\otimes t}\pi(y)^{\otimes t}\right]$ 
in Eq.~(\ref{predesign}), it can be easily shown that Lemma \ref{lemsym} and Theorem \ref{welchthm} (below) 
allows the following equivalent definition of a complex projective $t$-design.

\begin{dfn}\label{designdfn}
A finite set $\mathscr{D}\subset\C P^{d-1}$ is called a {\em $t$-design (of dimension $d$)\/} if 
\begin{equation}
\frac{1}{|\mathscr{D}|}\sum_{x\in\mathscr{D}}\,\pi(x)^{\otimes t}\;=\;\tbinom{d+t-1}{t}^{-1}\,\Pisymt\;.
\end{equation}
\end{dfn}
 
Seymour and Zaslavsky have shown that $t$-designs in $\C P^{d-1}$ exist for any $t$ and $d$~\cite{seymour}. It is 
necessary, however, that the number of design points satisfy~\cite{hoggar,bannai,dunkl,levenshtein05}
\begin{equation}
|\mathscr{D}|\;\geq\;\binom{d+\lceil t/2 \rceil -1}{\lceil t/2 \rceil}\binom{d+\lfloor t/2\rfloor -1}{\lfloor t/2\rfloor}\;.
\label{designbound}\end{equation}
A design which achieves this bound is called {\em tight\/}. Tight $t$-designs in $\C P^1$ are equivalent to tight 
spherical $t$-designs on the Euclidean 2-sphere [via Eq.~(\ref{embedding}) in Sec.~\ref{tightsec}]. Such designs exist only for 
$t=1,2,3,5$ (see e.g.~\cite{hardin}). When $d\geq 3$ it is known that tight $t$-designs in 
$\C P^{d-1}$ exist only for $t=1,2,3$~\cite{bannai,bannai2,hoggar25}. It is trivial that tight 1-designs exist in all dimensions.
Tight 2-designs have been conjectured to also exist for all $d$~\cite{zauner,renes}. Analytical constructions, however, are known 
only for $d\leq 10$ and $d=12,13,19$~\cite{zauner,renes,hoggar4,grassl,appleby,grassl2}. Examples of tight 3-designs are known only for $d=2,4,6$~\cite{hoggar}. 
When $d\geq 3$ and $t\geq 5$ the above bound can be improved by more than one~\cite{nikova,boyvalenkov,boyvalenkov2}. 

The concept of $t$-designs has been generalized to that of {\em weighted $t$-designs\/}~\cite{levenshtein,levenshtein2}. 
Each design point $x\in\mathscr{D}$ is then appointed a positive weight $w(x)$ under the normalization constraint
$\sum_{x\in\mathscr{D}}w(x)=1$. A countable set $\mathscr{S}$ endowed with a normalized weight function 
$w:\mathscr{S}\rightarrow[0,1]$ will be called a {\em weighted set\/} and denoted by the 
pair $(\mathscr{S},w)$. 

\begin{dfn}\label{wdesigndfn}
A finite weighted set $(\mathscr{D},w)$, $\mathscr{D}\subset\C P^{d-1}$, is called a {\em weighted $t$-design (of dimension $d$)\/} if 
\begin{equation}
\sum_{x\in\mathscr{D}}\,w(x)\pi(x)^{\otimes t}\;=\;\tbinom{d+t-1}{t}^{-1}\,\Pisymt\;.
\label{weighteddesign}\end{equation}
\end{dfn}

The weighted $t$-designs obviously incorporate the ``unweighted'' $t$-designs as the special case 
$w\equiv 1/|\mathscr{D}|$. Note that the normalization of $w$ is implied by the trace of Eq.~(\ref{weighteddesign}).
If we instead ``trace out'' only one subsystem of these $t$-partite operators, we can immediately deduce that every weighted 
$t$-design is also a weighted $(t-1)$-design. A weighted 1-design is known as a {\em tight (vector) frame\/} in the context of frame 
theory~\cite{christensen,daubechies2,casazza}, in which case the unnormalized states $\ket{\widetilde{x}}\equiv\sqrt{w(x)d}\,\ket{x}$ 
are the frame vectors, and Eq.~(\ref{weighteddesign}) is the tight frame condition: $\sum_{x\in\mathscr{D}}\ketbra{\widetilde{x}}=I$. 
In this form it is immediately apparent that we must have $|\mathscr{D}|\geq d$ for a weighted 1-design, with equality only 
if the frame vectors $\ket{\widetilde{x}}$ form an orthonormal basis for $\C^d$. 
The 2-design case is treated in the following theorem.

\begin{thm}\label{2designthm}
Let $(\mathscr{D},w)$ be a weighted $2$-design of dimension $d$. Then $|\mathscr{D}|\geq d^2$ 
with equality only if $w\equiv 1/|\mathscr{D}|$ and $|\braket{x}{y}|^2 = 1/(d+1)$ for all $x,y\in\mathscr{D}$ with $x\neq y$. 
\end{thm}
\begin{proof}
By the definition of a weighted 2-design, 
\begin{equation}
\sum_{x\in\mathscr{D}}\,w(x)\pi(x)\otimes\pi(x)\;=\;\frac{2}{d(d+1)}\,\Pisym^{(2)}\;=\;\frac{1}{d(d+1)}\,\sum_{j,k}\ketbra{e_j}\otimes\ketbra{e_k}+\ket{e_j}\bra{e_k}\otimes\ket{e_k}\bra{e_j}\;, 
\end{equation}
where $\{\ket{e_k}\}_{k=1}^d$ is an orthonormal basis for $\C^d$. Now if we multiply both sides of this equation by 
$A\otimes I$, where $A$ is an arbitrary linear operator, and then trace out the first subsystem, we find that
\begin{eqnarray}
\sum_{x\in\mathscr{D}}\,w(x)\tr[\pi(x)A]\pi(x) &=& \frac{1}{d(d+1)}\,\sum_{j,k}\bra{e_j}A\ket{e_j}\ketbra{e_k}+\ket{e_k}\bra{e_k}A\ket{e_j}\bra{e_j} \\
&=& \frac{1}{d(d+1)}\,\Big(\tr(A)I+A\Big)
\end{eqnarray}
and thus any $A\in\End(\C^d)$ can be rewritten as a linear combination of the design projectors:
\begin{equation}
A\;=\;d\sum_{x\in\mathscr{D}}w(x)\Big((d+1)\tr[\pi(x)A]-\tr(A)\Big)\pi(x) 
\label{2designprf1}\end{equation}
where we have used the fact that a 2-design is also a 1-design, i.e. $I=d\sum_{x\in\mathscr{D}}w(x)\pi(x)$. 
Consequently, the design projectors $\pi(x)$ span $\End(\C^d)\cong\C^{d^2}$, and thus, there must be at least
$d^2$ many. Furthermore, when $|\mathscr{D}|=d^2$ these operators must be linearly independent. Assuming this 
to be the case, and choosing $A=\pi(y)$ in Eq.~(\ref{2designprf1}), for some fixed $y\in\mathscr{D}$, we find that
\begin{equation}
\big(w(y)d^2-1\big)\pi(y)\;+\;d\sum_{x\neq y}w(x)\Big((d+1)\tr[\pi(x)\pi(y)]-1\Big)\pi(x) \;=\; 0\;,
\end{equation}
which, given the linear independence of the design projectors, can be satisfied only if $w(y)=1/d^2=1/|\mathscr{D}|$ and
$\tr[\pi(x)\pi(y)]=|\braket{x}{y}|^2 = 1/(d+1)$ for all $x\neq y$. The same is true for all $y\in\mathscr{D}$.
\end{proof}

Theorem~\ref{2designthm} is essentially a special case of the results of Levenshtein~\cite{levenshtein,levenshtein2}.    
In fact, the above lower bound [Eq.~(\ref{designbound})] also holds for weighted $t$-designs, with equality occurring 
only if the design has uniform weight, i.e. $w\equiv 1/|\mathscr{D}|$. The current proof, however, takes a 
form which incorporates the theme of this article. Like in the specific 2-design case, more can be said about the 
structure of $t$-designs when Eq.~(\ref{designbound}) is satisfied with equality. Our interest lies 
only with the 2-designs, however, and thus we defer further results in this direction to the work of Bannai and Hoggar~\cite{hoggar,hoggar2,hoggar25,hoggar3,bannai,bannai2}.

We have introduced complex projective $t$-designs as a special type of weighted subset of 
$\C P^{d-1}$. Notice that the weight function of an arbitrary weighted set $(\mathscr{S},w)$ may be trivially 
extended to a countably additive measure on the power set $2^\mathscr{S}$. We will use this observation to 
generalize the concept of $t$-designs one step further. Let $\mathfrak{B}(\mathscr{S})$ denote the Borel $\sigma$-algebra of 
$\mathscr{S}$. In the following situation, a set $\mathscr{S}$ endowed with a probability measure 
$\omega:\mathfrak{B}(\mathscr{S})\rightarrow[0,1]$, i.e. a (Borel) probability space, will be called a {\em distribution\/} and 
denoted by the pair $(\mathscr{S},\omega)$.

\begin{dfn}\label{gdesigndfn}
A distribution $(\mathscr{D},\omega)$, $\mathscr{D}\subseteq\C P^{d-1}$, is called a {\em generalized $t$-design (of dimension $d$)\/} if 
\begin{equation}\label{design}
\int_\mathscr{D} \d\omega(x)\,\pi(x)^{\otimes t}\;=\;\tbinom{d+t-1}{t}^{-1}\,\Pisymt\;.
\end{equation}
\end{dfn}

In this definition the Lebesgue-Stieltjes integral is used, which reduces to a discrete sum when 
$\mathscr{D}$ is countable. A generalized $t$-design is thus a weighted $t$-design when $\mathscr{D}$ is a finite 
set. Again, every generalized $t$-design is also a generalized $(t-1)$-design, and by Lemma~\ref{lemsym}, 
$(\C P^{d-1},\muu)$ is a generalized $t$-design for all $t$. 

By allowing any distribution of points in $\C P^{d-1}$ 
which satisfies Eq.~(\ref{design}) to be called a ``generalized'' $t$-design, we have in fact contradicted an important 
purpose of designs, which is to convert integrals into finite sums. In this article, however, we will allow this discrepancy and 
henceforth refer to both weighted and ``generalized'' complex projective $t$-designs as simply $t$-designs.
The task of finding $t$-designs is facilitated by the following theorem (see e.g.~\cite{levenshtein,konig}). 

\begin{thm}\label{welchthm}
Let $(\mathscr{S},\omega)$, $\mathscr{S}\subseteq\C P^{d-1}$, be a distribution. Then for any $t\geq 1$,
\begin{equation}\label{welchbound}
\iint_\mathscr{S}\d\omega(x)\d\omega(y)\,|\braket{x}{y}|^{2t}\;\geq\;\tbinom{d+t-1}{t}^{-1}\;,
\end{equation}
with equality iff $(\mathscr{S},\omega)$ is a $t$-design.
\end{thm}
\begin{proof}
Consider an arbitrary distribution $(\mathscr{S},\omega)$ and define 
\begin{equation}
S\;\equiv\;\int_{\mathscr{S}}\d\omega(x)\,\pi(x)^{\otimes t}
\end{equation}
which has support only on the totally symmetric subspace of $(\C^d)^{\otimes t}$. This positive operator 
can thus have at most $\dsym=\tbinom{d+t-1}{t}$ nonzero eigenvalues $\lambda_1,\dots,\lambda_{\dsym}$, which 
satisfy the equations
\begin{equation}
\tr(S) \;=\; \int_{\mathscr{S}}\d\omega(x) \;=\; 1 \;=\; \sum_{k=1}^{\dsym} \lambda_k\;, \;\qquad\text{and}\qquad\;  \tr({S}^2) \;=\; \iint_{\mathscr{S}} \d\omega(x)\d\omega(y) \, |\braket{x}{y}|^{2t} \;=\; \sum_{k=1}^{\dsym} {\lambda_k}^2\;.
\end{equation}
The lower bound [Eq.~(\ref{welchbound})] is apparent from the RHS of these equations. Under the normalization 
constraint expressed by the first, the second is bounded below: 
$\tr({S}^2)\geq 1/{\dsym}$. Equality can occur if and only if 
$\lambda_k=1/{\dsym}$ for all $k$, or equivalently $S=\Pisymt/{\dsym}$, which is the defining 
property of a $t$-design.\end{proof}

This theorem allows us to check whether a distribution of points in $\C P^{d-1}$ forms a $t$-design by considering only the angles 
between the supposed design elements. It also shows that $t$-designs can be found numerically by 
parametrizing a distribution and minimizing the LHS of Eq.~(\ref{welchbound}). The lower bound is in 
fact a straightforward generalization of the Welch bound~\cite{welch}.

\section{Operator frames}
\label{framesec}

Frame theory~\cite{christensen,daubechies2,casazza} provides a natural setting for the study of informationally complete 
quantum measurements~\cite{dariano}. In this section we will introduce some of the important concepts of this theory
that are relevant to the current investigation. Before beginning, however, we will need to introduce the concept of a superoperator.

Following Caves~\cite{caves} we will write a linear operator $A$ in vector notation as $\Ket{A}$. The vector space of all such operators, 
$\End(\C^d)\cong\C^{d^2}$, equipped with the Hilbert-Schmidt inner product $\BraKet{A}{B}\equiv\tr(A^\dag B)$, 
is a Hilbert space, where we think of $\Bra{A}$ as an operator ``bra'' and $\Ket{B}$ as an operator ``ket.''
Addition and scalar multiplication of operator kets then follows that for operators, e.g. $a\Ket{A}+b\Ket{B}=\Ket{aA+bB}$.
The usefulness of this notation becomes apparent when we consider linear maps on operators, i.e. superoperators. 
Given an orthonormal operator basis $\{E_k\}_{k=1}^{d^2}\subset\End(\C^d)$, $\BraKet{E_j}{E_k}=\delta(j,k)$, 
a superoperator $\mathcal{S}\in\End(\End(\C^d))\cong\C^{d^4}$ may be written in two different ways:
\begin{equation}
\mathcal{S}\;=\;\sum_{j,k}s_{jk}\,E_j\odot{E_k}^\dag\;=\;\sum_{j,k}s_{jk}\,\Ket{E_j}\Bra{E_k} \qquad\quad (s_{jk}\in\C)\;.
\end{equation}
The first representation illustrates the {\em ordinary\/} action of the superoperator, 
\begin{equation}
\mathcal{S}(A)\;\equiv\;\sum_{j,k}s_{jk}E_j A{E_k}^\dag\;,
\end{equation} 
which amounts to inserting $A$ into the location of the `$\odot$' symbol. The second reflects the {\em left-right\/} 
action, 
\begin{equation}
\mathcal{S}\Ket{A}\;\equiv\;\sum_{j,k}s_{jk}\Ket{E_j}\BraKet{E_k}{A}\;=\;\sum_{j,k}s_{jk}{E_j}\tr\big({E_k}^\dag{A}\big)\;,
\end{equation} 
where the superoperator acts on operators just like an operator on vectors. It is this second ``non-standard'' 
action which will be useful in the current context. The identity superoperators relative to the ordinary and
left-right actions are, respectively, $\mathcal{I}\equiv I\odot I$ and $\Is\equiv\sum_k\Ket{E_k}\Bra{E_k}$.
Further results on superoperators in the current notation can be found in Ref.'s~\cite{rungta,rungta2}. 

The notion of an informationally complete POVM is naturally related to that of a frame, or more specifically, an 
``operator'' frame. Frames generalize the notion of bases. We call a countable family of operators 
$\{A(x)\}_{x\in\mathscr{X}}\subset\End(\C^d)$ an {\em operator frame\/} if there exist constants 
$0<a\leq b<\infty$ such that 
\begin{equation}
a\BraKet{C}{C} \;\leq\; \sum_{x\in\mathscr{X}}\big|\BraKetb{A(x)}{C}\big|^2 \;\leq\; b\BraKet{C}{C}
\label{framecondition}\end{equation}
for all $C\in\End(\C^d)$. For example, all finite linearly spanning subsets of $\End(\C^d)$ are operator frames. 
When $a=b$ the frame is called {\em tight\/}~\cite{daubechies}. Tight frames are those frames which are most like 
orthonormal bases (see e.g.~\cite{casazza2}). An operator frame with cardinality $|\mathscr{X}|=d^2$, i.e. an operator 
basis, is tight if and only if it is an orthonormal basis. For every frame $\{A(x)\}_{x\in\mathscr{X}}$ there is 
a {\em dual frame\/} $\{B(x)\}_{x\in\mathscr{X}}$, such that 
\begin{equation}
\sum_{x\in\mathscr{X}} \Ketb{B(x)}\Brab{A(x)} \; =\; \Is\;, 
\end{equation}
and hence, 
\begin{equation}
C \;=\; \sum_{x\in\mathscr{X}}\BraKetb{A(x)}{C}B(x) \;=\; \sum_{x\in\mathscr{X}}\BraKetb{B(x)}{C}A(x)
\end{equation} 
for all $C\in\End(\C^d)$. Although when $|\mathscr{X}|>d^2$ there are different choices for the dual frame~\cite{li}, the most ``economical'' choice 
(see Proposition~3.2.4 of \cite{daubechies2}) is the {\em canonical dual frame\/} $\{\tilde{A}(x)\}_{x\in\mathscr{X}}$, 
\begin{equation}
\Ketb{\tilde{A}(x)} \;\equiv\; \mathcal{A}^{-1}\Ketb{A(x)}\;,
\label{dualframe}\end{equation}
where the {\em frame superoperator\/} 
\begin{equation}
\mathcal{A} \;\equiv\; \sum_{x\in\mathscr{X}}\KetBrab{A(x)}\;, 
\end{equation}
so that 
\begin{equation}
\sum_{x\in\mathscr{X}} \Ketb{\tilde{A}(x)}\Brab{A(x)} \;=\; \sum_{x\in\mathscr{X}} \mathcal{A}^{-1}\Ketb{A(x)}\Brab{A(x)} \;=\; \mathcal{A}^{-1}\mathcal{A} \;=\; \Is
\end{equation}
as required. Note that the inverse of $\mathcal{A}$ is taken with respect to left-right action, and exists 
whenever $\{A(x)\}_{x\in\mathscr{X}}$ is an operator frame. 
A tight operator frame is one with $\mathcal{A}=a\Is$, and thus trivially 
$\Ketb{\tilde{A}(x)}=\Ketb{A(x)}/a$. In general, however, inverting the frame superoperator will be a 
difficult analytical task. 

In this article we prefer the concept of generalized (or ``continuous'') frames~\cite{ali,kaiser,christensen} over the 
preceding more common notion. Suppose now that the set $\mathscr{X}$ 
(which need no longer be countable) is endowed with a positive measure $\alpha:\mathfrak{B}(\mathscr{X})\rightarrow [0,\infty]$. We call a family of operators 
$\{A(x)\}_{x\in\mathscr{X}}\subseteq\End(\C^d)$ a {\em generalized operator frame (with respect to $\alpha$)\/} if 
there exist constants $0<a\leq b<\infty$ such that 
\begin{equation}
a\BraKet{C}{C} \;\leq\; \int_{\mathscr{X}}\d\alpha(x)\,\big|\BraKetb{A(x)}{C}\big|^2 \;\leq\; b\BraKet{C}{C}
\label{frameconditionc}\end{equation}
for all $C\in\End(\C^d)$. This definition reduces to the above discrete case when $\mathscr{X}$ is countable and 
$\alpha$ is the counting measure. Again, for every frame $\{A(x)\}_{x\in\mathscr{X}}$ there is a dual frame 
$\{B(x)\}_{x\in\mathscr{X}}$ such that 
\begin{equation}
\int_{\mathscr{X}}\d\alpha(x)\,\Ketb{B(x)}\Brab{A(x)} \; =\; \Is\;, 
\label{frameresolution}\end{equation}
and the canonical dual frame $\{\tilde{A}(x)\}_{x\in\mathscr{X}}$ is defined through Eq.~(\ref{dualframe}), where now the frame 
superoperator 
\begin{equation}
\mathcal{A} \;\equiv\; \int_{\mathscr{X}}\d\alpha(x)\KetBrab{A(x)} \;.
\label{framesuper}\end{equation}
A generalized tight operator frame is also defined in analogy to the discrete case. 
\begin{dfn}\label{tightframedfn}
An operator frame $\{A(x)\}_{x\in\mathscr{X}}\subseteq\End(\C^d)$ with respect to the measure $\alpha$ is called {\em tight\/} if 
\begin{equation}
\int_{\mathscr{X}}\d\alpha(x)\,\Ketb{A(x)}\Brab{A(x)} \; =\; a\Is \;,
\label{tightframe}\end{equation}
for some constant $a>0$, i.e. $\mathcal{A}=a\Is$.
\end{dfn}

The argument that tight frames are ``as close as possible'' to orthonormal bases comes from this resolution of unity 
[Eq.~(\ref{tightframe})] and the following inequality (see e.g.~\cite{casazza2}), which is called the frame 
bound. Let `$\Tr$' denote the superoperator trace.

\begin{thm}\label{framethm}
Let $\{A(x)\}_{x\in\mathscr{X}}\subseteq\End(\C^d)$ be an operator frame with respect to the measure $\alpha$. Then
\begin{equation}
\iint_{\mathscr{X}}\d\alpha(x)\d\alpha(y)\,\big|\BraKetb{A(x)}{A(y)}\big|^2 \;\geq\; \frac{\big(\Tr(\mathcal{A})\big)^2}{d^2} \;,
\label{framebound}\end{equation}
with equality if and only if $\{A(x)\}_{x\in\mathscr{X}}$ is a tight operator frame.
\end{thm}
\begin{proof}
Let $\lambda_1,\dots,\lambda_{d^2}>0$ denote the left-right eigenvalues of $\mathcal{A}$. The LHS of Eq.~(\ref{framebound}) can be 
rewritten as
\begin{equation}
\Tr(\mathcal{A}^2) \;=\; \sum_{k=1}^{d^2} {\lambda_k}^2 \;, 
\end{equation}   
which under the constraint $\sum_{k=1}^{d^2}\lambda_k=\Tr(\mathcal{A})$ takes its minimum value if and only if 
$\lambda_1=\dots=\lambda_{d^2}=\Tr(\mathcal{A})/d^2$, i.e. $\mathcal{A}=(\Tr(\mathcal{A})/d^2)\Is$, which means 
$\{A(x)\}_{x\in\mathscr{X}}$ is a tight operator frame with $a=\Tr(\mathcal{A})/d^2$. The minimum is $a^2d^2$.
\end{proof}

The frame bound can be considered a variant of the Welch bound (Theorem~\ref{welchthm}) with $t=1$. 
Theorem~\ref{framethm} shows that tight frames are those which minimize the average correlation amongst the frame elements. 
It is straightforward to show that tight operator frames exist for all $d$ and $|\mathscr{X}|\geq d^2$. 
Given any operator frame $\{A(x)\}_{x\in\mathscr{X}}$ we can construct a tight frame through the use of the frame 
superoperator, e.g. $\{\mathcal{A}^{-1/2}\Ket{A(x)}\}_{x\in\mathscr{X}}$. 
In fact, explicit examples of tight {\em unitary\/} operator frames are known for all $d$ and $|\mathscr{X}|\geq d^2$~\cite{horn}. 
Unitary 2-designs have also recently been considered~\cite{dankert} (unitary 1-designs are equivalent to tight unitary operator frames). 
These examples are important for the implementation of certain types of quantum processes, which are defined by rewriting 
Eq.~(\ref{tightframe}) (and its $t$-design generalization) in terms of the ordinary action of a superoperator. 
For example, tight unitary operator frames implement the depolarizing channel. To relate the concept of tight frames to 
informationally complete POVMs we need to instead consider frames on a subspace of $\End(\C^d)$. This will be 
done in Sec.~\ref{tightsec}.

\section{Informationally complete quantum measurements}
\label{icpovmsec}

The outcome statistics of a quantum measurement are described by a positive-operator-valued measure (POVM) 
\cite{davies,holevo,kraus,busch2}. That is, an operator-valued function defined on a $\sigma$-algebra over 
the set $\mathscr{X}$ of outcomes, $F:\mathfrak{B}(\mathscr{X})\rightarrow\End(\C^d)$, which satisfies
(1) $F(\mathscr{S})\geq 0$ for all $\mathscr{S}\in\mathfrak{B}(\mathscr{X})$ with equality if 
$\mathscr{S}=\emptyset$, (2)~$F(\bigcup_{k=1}^\infty\mathscr{S}_k)=\sum_{k=1}^\infty F(\mathscr{S}_k)$ for any 
sequence of disjoint sets $\mathscr{S}_k\in\mathfrak{B}(\mathscr{X})$, and (3) the normalization constraint 
$F(\mathscr{X})=I$. In this article we always take $\mathfrak{B}(\mathscr{X})$ to be the Borel $\sigma$-algebra.

An informationally complete quantum measurement~\cite{prugovecki} is one with the property that 
each quantum state $\rho\in\Qd\equiv \big\{A\in\End(\C^d)\,|\,A\geq 0\,,\,\tr(A)=1\big\}$ is uniquely determined by its 
measurement statistics $p(\mathscr{S})\equiv\tr\left[F(\mathscr{S})\rho\right]$. 
Consequently, given multiple copies of a system in an unknown state, a sequence of measurements will give an estimate of the 
statistics, and hence, identify the state. The measure 
$F$ is then called an {\em informationally complete POVM (IC-POVM)\/}. 

\begin{dfn}
A POVM $F:\mathfrak{B}(\mathscr{X})\rightarrow\End(\C^d)$ is called {\em informationally complete\/} if 
for each pair of distinct quantum states $\rho\neq\sigma\in\Qd$ there exists an event 
$\mathscr{S}\in\mathfrak{B}(\mathscr{X})$ such that 
$\tr\left[F(\mathscr{S})\rho\right]\neq\tr\left[F(\mathscr{S})\sigma\right]$.  
\end{dfn}

When a quantum measurement has a countable number of outcomes, the indexed set of POVM elements
$\{F(x)\}_{x\in\mathscr{X}}$ completely characterizes $F$, and is thus often referred to as the 
``POVM.'' We will call such measurements {\em discrete\/}, or {\em finite\/} if we additionally have 
$|\mathscr{X}|<\infty$. A discrete POVM is informationally complete if 
and only if for each pair of distinct quantum states $\rho\neq\sigma\in\Qd$ there exists an outcome 
$x\in\mathscr{X}$ such that $\tr\left[F(x)\rho\right]\neq\tr\left[F(x)\sigma\right]$. 

To show how a quantum state can be reconstructed from its measurement statistics, we will first need to express 
$F$ in a standard form. Consider an arbitrary quantum measurement. The POVM defines a natural real-valued 
{\em trace measure\/}~\cite{rosenberg}, $\tau(\mathscr{S})\equiv\tr[F(\mathscr{S})]$, which inherits the 
normalization $\tau(\mathscr{X})=d$. Since each matrix element of $F$ is a complex valued measure which is 
absolutely continuous with respect to the nonnegative finite measure $\tau$, the POVM can be expressed as
\begin{equation}
F(\mathscr{S}) \;=\; \int_\mathscr{S}\d\tau(x)\, \Ft(x) \;\equiv\; \int_\mathscr{S}\d\tau(x)\, \Pt(x)\;,
\label{POVDeq}\end{equation}
where the Radon-Nikodym derivative $\Ft:\mathscr{X}\rightarrow\End(\C^d)$ is a positive-operator-valued density 
(POVD) which is uniquely defined up to a set of zero $\tau$-measure. We will set $\Ft\equiv\Pt$. Note that our 
choice of scalar measure implies that $\tr(\Pt)=1$, $\tau$-almost everywhere. When $\Pt$ also has unit rank we 
call $F$ a {\em rank-one POVM\/}, in which case it is natural to have $\mathscr{X}\subseteq\C P^{d-1}$ and 
then $\Pt\equiv\pi$. A rank-one POVM is of course equivalent to a 1-design, i.e. a tight vector frame~\cite{eldar}.
In the special case of a discrete quantum measurement the Radon-Nikodym derivative is simply $\Pt(x)\equiv\Ft(x)=F(x)/\tr[F(x)]$. 

For an arbitrary POVM $F$, define the superoperator
\begin{equation}
\FFt\;\equiv\; \int_\mathscr{X} \d\tau(x)\,\KetBrab{\Pt(x)}\;,
\label{framesuperoperator}\end{equation}
which is positive and bounded under the left-right action:
\begin{equation}
0\;\leq\;\Bra{A}\FFt\Ket{A}\;=\; \int_\mathscr{X} \d\tau(x)\,\big|\BraKetb{A}{\Pt(x)}\big|^2 \;\leq\; \int_\mathscr{X} \d\tau(x)\,\BraKetb{P(x)}{P(x)}\BraKet{A}{A} \;\leq\; \int_\mathscr{X} \d\tau(x)\,\BraKet{A}{A}\;=\;  d\BraKet{A}{A}
\end{equation} 
for all $A\in\End(\C^d)$, where we have used the Cauchy-Schwarz inequality and then the fact that $\tr(P^2)\leq 1$. 
Now consider the following straightforward result. 

\begin{prp}
Let $F:\mathfrak{B}(\mathscr{X})\rightarrow\End(\C^d)$ be a POVM. Then $F$ is informationally complete iff there 
exists a constant $a>0$ such that $\Bra{A}\FFt\Ket{A}\geq a\BraKet{A}{A}$ for all $A\in\End(\C^d)$.
\label{icpovmprp}\end{prp} 

\begin{proof}
Suppose $F$ is informationally complete. If there existed an operator $A\neq 0$ such that 
\begin{equation}
\Bra{A}\FFt\Ket{A}\;=\;\int_\mathscr{X}\d\tau(x)\,\big|\!\tr[\Pt(x)A]\big|^2\;=\;0\;,
\end{equation}
then we must have $\tr(\Pt A)=0$, $\tau$-almost everywhere. This operator must therefore be traceless:
\begin{equation}
\tr(A)\;=\;\tr[F(\mathscr{X})A]\;=\;\int_\mathscr{X}\d\tau(x)\,\tr[\Pt(x)A]\;=\;0\;.
\end{equation}
Now for any state $\rho\in\Qd$ we can define the state $\sigma=\rho+\epsilon(A+A^\dag)$, where $\epsilon>0$ 
is chosen small enough such that $\sigma\geq 0$. Then 
\begin{equation}
\tr[F(\mathscr{S})\sigma]\;=\;\tr[F(\mathscr{S})\rho]+\epsilon\int_\mathscr{S}\d\tau(x)\,\Big(\tr[\Pt(x)A]+\tr[\Pt(x)A]^* \Big) \;=\; \tr[F(\mathscr{S})\rho]
\end{equation}
for all $\mathscr{S}\in\mathfrak{B}(\mathscr{X})$, with $\sigma\neq\rho$. This means $F$ could not 
have been informationally complete. Thus for IC-POVMs, $\FFt$ will always be strictly positive relative to the 
left-right action. The converse is also true. If for the distinct quantum states 
$\rho\neq\sigma\in\Qd$ we have
\begin{equation}
\Bra{\rho-\sigma}\FFt\Ket{\rho-\sigma}\;=\;\int_\mathscr{X}\d\tau(x)\,\big|\!\tr[\Pt(x)(\rho-\sigma)]\big|^2 \;>\;0 
\end{equation}
then there must exist an event $\mathscr{S}\in\mathfrak{B}(\mathscr{X})$, such that   
\begin{equation}
\int_\mathscr{S}\d\tau(x)\,\tr[\Pt(x)(\rho-\sigma)] \;\neq\;0\;, 
\end{equation}
or equivalently, $\tr[F(\mathscr{S})\rho]\neq\tr[F(\mathscr{S})\sigma]$, which means $F$ is informationally complete.
\end{proof}

Note that the proof of Proposition~\ref{icpovmprp} made no reference to our particular choice of scalar measure.
We could also express the POVM in terms of another. However the trace measure guarantees the boundedness of 
the superoperator $\mathcal{F}$ and was found to be the best choice for a canonical scalar measure in the 
current context. 

When a POVM $F$ is informationally complete, in which case we have just shown that the corresponding 
superoperator $\FFt$ has full rank relative to the left-right action, the POVD $\Pt$ can be considered a generalized operator frame 
with respect to $\tau$. The canonical dual frame then defines a {\em reconstruction operator-valued density\/}  
\begin{equation}
\Ket{\Rt} \;\equiv\; {\FFt}^{-1}\Ket{\Pt}\;,
\label{reconstructOVD}\end{equation}
where the inverse of $\FFt$, which we now call the {\em POVM superoperator}, is taken with respect to the left-right action. 
The identity
\begin{equation}
\int_\mathscr{X}\d\tau(x)\,\Ketb{\Rt(x)}\Brab{\Pt(x)} \;=\; \int_\mathscr{X}\d\tau(x)\,{\FFt}^{-1}\KetBrab{\Pt(x)} \;=\;{\FFt}^{-1}\FFt\;=\;\Is \;,
\label{prereconstruct}\end{equation}
then allows state reconstruction in terms of the measurement statistics:
\begin{equation}
\rho \;=\; \int_\mathscr{X}\d\tau(x)\, \tr[\Pt(x)\rho]\Rt(x) \;=\; \int_\mathscr{X}\tr[\d F(x)\rho]\Rt(x) \;=\; \int_\mathscr{X}\d p(x)\Rt(x)\;.
\label{reconstruct}\end{equation}
where $p(\mathscr{S})\equiv\tr\left[F(\mathscr{S})\rho\right]=\int_\mathscr{S}\d\tau(x)\tr[\Pt(x)\rho]$. This 
{\em state-reconstruction formula\/} is an immediate consequence of the left-right action of Eq.~(\ref{prereconstruct}) on $\Ket{\rho}$.

We will now give some useful properties of the reconstruction operator-valued density (OVD) which will be needed later 
in the article. Although $\Rt$ is generally not positive, it inherits all other properties of $\Pt$.  
For example, we know that $\Rt$ is Hermitian since $\FFt$, and thus ${\FFt}^{-1}$, maps Hermitian operators to 
Hermitian operators. Additionally, the left-right action of Eq.~(\ref{prereconstruct}) on $\Ket{I}$ shows that 
\begin{equation}
\int_\mathscr{X}\d\tau(x)\Rt(x) \;=\; I\;.
\end{equation}
Notice that for an arbitrary POVM, the identity operator is always a left-right 
eigenvector of the POVM superoperator:
\begin{equation}
\FFt\Ket{I}\;=\; \int_\mathscr{X} \d\tau(x)\,\Ketb{\Pt(x)}\BraKetb{\Pt(x)}{I}\;=\; \int_\mathscr{X} \d\tau(x)\,\Ketb{\Pt(x)}\;=\; \int_\mathscr{X}\,\Ketb{\d F(x)}\;=\;\Ket{I}\;,
\label{identityeigenvector}\end{equation}
using $\tr(\Pt)=1$ and the normalization of the POVM. Thus $\Ket{I}$ is also an eigenvector of ${\FFt}^{-1}$, and we obtain
\begin{equation}
\tr(\Rt)\;=\; \BraKet{I}{\Rt}\;=\;\Bra{I}{\FFt}^{-1}\Ket{\Pt}\;=\;\BraKet{I}{\Pt}\;=\;\tr(\Pt)\;=\;1 \;.
\end{equation}
Finally, it is straightforward to confirm that
\begin{equation}
{\FFt}^{-1}\;=\; \int_\mathscr{X} \d\tau(x)\,\KetBrab{\Rt(x)}\;.
\label{framesuperoperatorinverse}\end{equation}

Note that we need $|\mathscr{X}|\geq d^2$ 
for $F$ to be informationally complete. If this were not the case then $\FFt$ could not have full rank. An IC-POVM 
with $|\mathscr{X}|=d^2$ is called {\em minimal\/}. In this case the reconstruction OVD is unique. 
In general, however, there will be many different choices. When $F$ is a discrete IC-POVM
the trace measure can be replaced by the counting measure~\cite{dariano} in Eq.~(\ref{POVDeq}). Then $P'(x)=F(x)$, the POVM 
superoperator is $\mathcal{F}'=\sum_{x\in\mathscr{X}}\KetBrab{F(x)}$, and $\Ket{\Rt'(x)}=\FFt'^{\,-1}\Ket{F(x)}$ say. 
In this case we also have $\rho=\sum_{x\in\mathscr{X}} p(x)\Rt'(x)$.
If it were not already obvious, it is now clear from the superoperator $\mathcal{F}'$ that $F$ is informationally complete if and only if 
$\{F(x)\}_{x\in\mathscr{X}}$ spans $\End(\C^d)$. Although the counting measure might seem more convenient, 
in Sec.~\ref{estimatesec} we will show that the canonical dual frame with respect to the trace measure is 
the optimal choice for quantum state tomography.

\section{Tight IC-POVMs}
\label{tightsec}

The notion of an informationally complete POVM is naturally related to that of an operator frame.
In the previous section we showed how to reconstruct a quantum state from its measurement statistics for an 
arbitrary IC-POVM. The procedure required inverting a superoperator, however, which may not be a straightforward 
analytical task. In this section we will investigate a class of IC-POVMs which share a particularly simple 
state-reconstruction formula. In analogy with a tight frame, these IC-POVMS will be called {\em tight IC-POVMs\/}.

Although pure states correspond to rays in a complex vector space, the most natural setting in which to study a general quantum 
state is Euclidean space. The set of all quantum states $\Qd$ is embedded in $\R^{d^2-1}$ as follows. Note that each $\rho\in\Qd$ may be associated with a traceless Hermitian operator under the mapping 
$\rho\rightarrow\rho-I/d$. Equipped with the Hilbert-Schmidt inner product 
$\BraKet{A}{B}\equiv\tr(A^\dag B)$, which induces the Frobenius norm $\|A\|\equiv\sqrt{\BraKet{A}{A}}$, the set of all 
traceless Hermitian operators $\Hd\equiv\{A\in\End(\C^d)\,|\,A^\dag=A\,,\,\tr(A)=0\}\cong\R^{d^2-1}$, forms a 
real inner product space in which the images of pure states lie on a sphere, $\|\pi(\psi)-I/d\|=\sqrt{(d-1)/d}$, and the 
images of mixed states within. In the special case $d=2$, this isometric embedding maps quantum states surjectively onto 
a ball in $\Htwo\cong\R^3$, realizing the Bloch-sphere representation of a qubit, but is otherwise only injective.

Let us now reconsider the POVM superoperator of an arbitrary POVM [Eq.~(\ref{framesuperoperator})] in this setting. It is straightforward to 
confirm that we have the decomposition 
\begin{equation}
\mathcal{F}\;=\;\frac{\mathcal{I}}{d}\;+\;\int_\mathscr{X} \d\tau(x)\,\KetBrab{\Pt(x)-I/d}\;.
\label{POVMdecomp}\end{equation}
The superoperator $\mathcal{I}/d=\KetBra{I}/d$ is in fact an eigenprojector [Eq.~(\ref{identityeigenvector})]. It left-right projects onto the subspace spanned by 
the identity, whose orthogonal complement, the $(d^2-1)$-dimensional subspace of traceless operators, is $\mathcal{F}$-invariant.
Define $\Ptr\equiv\Is-\mathcal{I}/d$, which left-right projects onto this latter subspace.  
The action of $\Ptr$ on a quantum state then realizes the above embedding into $\Hd$:
\begin{equation}
\Ptr\Ket{\rho} \;=\; \Ket{\rho-I/d} \;.
\label{embedding}\end{equation} 
Let $\Ih$ denote the identity superoperator for $\Hd$ under the left-right action. Noting that $\Pt-I/d$ is a traceless Hermitian OVD, i.e. 
$\Pt(x)-I/d\in\Hd$ for all $x\in\mathscr{X}$, we are now ready to define a tight IC-POVM.  

\begin{dfn}\label{tightdfn}
Let $F:\mathfrak{B}(\mathscr{X})\rightarrow\End(\C^d)$ be a POVM. Then $F$ is called a {\em tight IC-POVM\/} if the 
OVD $\Pt-I/d$ forms a tight operator frame (with respect to $\tau$) in $\Hd$, i.e.
\begin{equation}
\int_\mathscr{X} \d\tau(x)\,\KetBrab{\Pt(x)-I/d}\;=\;a\Ih\;,
\label{tightdfneq}\end{equation}
or equivalently, $\Ptr\mathcal{F}\Ptr=a\Ptr$, for some constant $a>0$. 
\end{dfn}
Tight IC-POVMs are precisely those POVMs whose images under $\Ptr$ form tight operator frames in $\Hd$. 
It is in this sense that they are claimed ``as close as possible'' to orthonormal bases for the 
space of quantum states. The constant $a$ can be found by taking the superoperator 
trace of Eq.~(\ref{tightdfneq}):
\begin{eqnarray}
a\;=\;a(F) &=& \frac{1}{d^2-1}\int_\mathscr{X} \d\tau(x)\,\BraKetb{\Pt(x)-I/d}{\Pt(x)-I/d} \\
&=& \frac{1}{d^2-1}\left(\int_\mathscr{X} \d\tau(x)\,\BraKetb{\Pt(x)}{\Pt(x)}\;-\;1\right)\;.
\end{eqnarray}
The POVM superoperator of a tight IC-POVM satisfies the identity 
\begin{equation}
\FFt\;=\;\frac{1}{d}\,\mathcal{I}\,+\,a\Ptr \;=\;a\Is\,+\,\frac{1-a}{d}\,\mathcal{I}\;.
\label{tightframesuperoperator}\end{equation} 
Since $a>0$ by definition, this superoperator obviously has full rank. Its inverse is 
\begin{equation}
{\FFt}^{-1}\;=\;\frac{1}{a}\,\Is\,-\,\frac{1-a}{ad}\,\mathcal{I}\;,
\end{equation} 
and thus the reconstruction OVD [Eq.~(\ref{reconstructOVD})] takes the form 
\begin{equation}
\Rt \;=\; \frac{1}{a}\,\Pt\,-\,\frac{1-a}{ad}\,I\;,
\label{tightreconstructionOVD}\end{equation}
where we have used the fact that $\tr(\Pt)=1$. A tight IC-POVM then has a particularly simple state-reconstruction formula [Eq.~(\ref{reconstruct})]:
\begin{equation}
\rho \;=\; \frac{1}{a}\int_\mathscr{X}\d p(x)\Pt(x)\;-\;\frac{1-a}{ad}\,I\;.
\label{reconstructtight}\end{equation}
This formula may also be derived without taking the inverse of the POVM superoperator, but by simply 
inspecting the left-right action of $\FFt$ on a quantum state under its definition 
[Eq.~(\ref{framesuperoperator})], and then under the above identity [Eq.~(\ref{tightframesuperoperator})].

The above formulae simplify further in the important special case of a tight {\em rank-one\/} IC-POVM. The frame 
constant then takes its maximum possible value:
\begin{equation}
a\;=\;a(F) \;=\; \frac{1}{d+1}\;.
\label{tightrank1ICPOVMa}\end{equation}
Since this is in fact {\em only\/} possible for rank-one POVMs, by noting that Eq.~(\ref{tightframesuperoperator}) can be taken as 
an alternative definition in the general case, we obtain the following elegant alternative definition of a tight rank-one IC-POVM. 
\begin{prp}\label{tight1prp}
Let $F:\mathfrak{B}(\mathscr{X})\rightarrow\End(\C^d)$ be a POVM. Then $F$ is a tight rank-one IC-POVM iff
\begin{equation}
\FFt\;=\;\frac{\Is+\mathcal{I}}{d+1}\;.
\label{tightframesuperoperator1}\end{equation}
\end{prp}
The state-reconstruction formula for a tight rank-one IC-POVM also takes an elegant form:
\begin{equation}
\rho \;=\; (d+1)\int_\mathscr{X}\d p(x)\pi(x)\;-\;I\;,
\label{reconstructtight1}\end{equation}
where we have set the POVD to a rank-one projector, $\Pt\equiv\pi$, to emphasize the fact that we are now dealing 
exclusively with rank-one POVMs. It is then appropriate to consider the measurement 
outcomes as points in complex projective space, $\mathscr{X}\subseteq\C P^{d-1}$.

We can say some more about the structure of tight rank-one IC-POVMs. Note that $\End(\End(\C^d))\cong\End(\C^d)\otimes\End(\C^d)$. 
The natural isomorphism which enables this relationship amounts to replacing each `$\odot$' by  `$\otimes$' for a superoperator written in 
terms of its ordinary action. Rewriting Eq.~(\ref{tightframesuperoperator1}) in terms of the ordinary action 
\begin{equation}
\int_\mathscr{X} \d\tau(x)\,\pi(x)\odot{\pi(x)}\;=\;\frac{1}{d+1}\bigg(\sum_k E_k\odot {E_k}^\dag \,+\, I\odot I\bigg)\;,
\end{equation}
we see that the condition for a tight rank-one IC-POVM is equivalent to
\begin{eqnarray}
\int_\mathscr{X} \d\tau(x)\,\pi(x)\otimes{\pi(x)} &=& \frac{1}{d+1}\bigg(\sum_k E_k\otimes {E_k}^\dag \,+\, I\otimes I\bigg) \\
&=& \frac{1}{d+1}\,\Big(T \,+\, I\otimes I\Big) \\
&=& \frac{2}{d+1}\,\Pisym^{(2)} \label{tightconderiv}
\end{eqnarray}
where the swap, $T\equiv\sum_{j,k}\ket{e_j}\bra{e_k}\otimes\ket{e_k}\bra{e_j}=\sum_k E_k\otimes {E_k}^\dag$, 
for any orthonormal operator basis. With $\omega=\tau/d$ in Definition~\ref{gdesigndfn},
we see that tight rank-one IC-POVMs are equivalent to complex projective 2-designs. 
A diligent reader might have predicted this outcome from the proof of Theorem~\ref{2designthm}. 
\begin{prp}\label{tight1prp2}
A rank-one POVM, $F:\mathfrak{B}(\mathscr{X})\rightarrow\End(\C^d)$, $\mathscr{X}\subseteq\C P^{d-1}$, $\Pt\equiv\pi$,
is a tight IC-POVM iff the outcome distribution $(\mathscr{X},\tau/d)$ is a 2-design, i.e.
\begin{equation}
\int_\mathscr{X} \d\tau(x)\,\pi(x)\otimes\pi(x)\;=\;\frac{2}{d+1}\,\Pisym^{(2)}\;.
\end{equation}
\end{prp}

By Theorem~\ref{2designthm}, there is essentially a unique {\em minimal\/} tight rank-one IC-POVM for each dimension, 
i.e. one with $|\mathscr{X}|=d^2$. This IC-POVM corresponds to a tight 2-design, which in the context of quantum 
measurements, is called a {\em symmetric IC-POVM (SIC-POVM)\/}~\cite{renes}. The defining properties are 
$\tau(x)\equiv 1/d$, and 
\begin{equation}
\BraKetb{\pi(x)}{\pi(y)}\;=\;|\braket{x}{y}|^2\;=\;\frac{d\delta(x,y)+1}{d+1}\;.
\end{equation}
Although analytical constructions are known only for $d\leq 10$ 
and $d=12,13,19$~\cite{zauner,renes,hoggar4,grassl,appleby,grassl2}, SIC-POVMs are conjectured to exist in all 
dimensions~\cite{zauner,renes} (see also \cite{klappenecker2,colin,wootters2,godsil,flammia2}). Embedded in $\Hd\cong\R^{d^2-1}$, the elements of a SIC-POVM correspond to 
the vertices of a regular simplex:
\begin{equation}
\frac{d}{d-1}\BraKetb{\pi(x)-I/d}{\pi(y)-I/d}\;=\;\frac{d^2\delta(x,y)-1}{d^2-1}\;.
\end{equation}
However not all simplices will correspond to a POVM. The factor of $d/(d-1)$ is the result of embedding $\Qd$ into the 
sphere of radius $\sqrt{(d-1)/d}$ in $\Hd$ rather than the unit sphere.

Following the terminology of frame theory, a finite tight rank-one IC-POVM will be called {\em uniform\/} when 
$\tau(x)\equiv d/|\mathscr{X}|$, or {\em equiangular\/}~\cite{strohmer} if we additionally have 
$\BraKetb{\pi(x)}{\pi(y)}=|\braket{x}{y}|^2=c$ for all $x\neq y\in\mathscr{X}$ and some constant $c$. SIC-POVMs are examples of 
equiangular tight rank-one IC-POVMs. In fact, these are the only POVMs of this type. To show this, first note that 
the Welch bound [Eq.~(\ref{welchbound})] is saturated for both 
$t=1$ and $t=2$ in the case of a tight rank-one IC-POVM. Equiangularity then implies that, respectively,
\begin{equation}
c\;=\;\frac{n-d}{d(n-1)} \qquad\text{and}\qquad c^2\;=\;\frac{2n-d(d+1)}{d(d+1)(n-1)}\;,
\end{equation}
where we have set $|\mathscr{X}|=n$. The only solution to these equations is $n=d^2$ and $c=1/(d+1)$. 

Another important example of a tight rank-one IC-POVM is a complete set of mutually unbiased bases (MUBs)~\cite{wootters,ivanovic}. 
That is, a set of $d+1$ orthonormal bases for $\C^d$ with a constant overlap of $1/d$ between elements of different bases:
\begin{equation}
\BraKetb{\pi(e_j^l)}{\pi(e_k^m)}\;=\;|\braket{e_j^l}{e_k^m}|^2\;=\;\left\{\begin{array}{ll}\delta(j,k)\,, &\; l=m \\  1/d\,, &\; l\neq m \end{array}\right.\;.
\label{MUBs}\end{equation}
Using Theorem~\ref{welchthm} it is straightforward to check that the union of $d+1$ MUBs $\mathscr{D}=\{e_k^m\,|\,1\leq k\leq d\,,\, 1\leq m\leq d+1\}$
forms a 2-design with uniform weight $w\equiv 1/|\mathscr{D}|=1/d(d+1)$ \cite{barnum,klappenecker}. Thus 
with $\tau(x)\equiv 1/(d+1)$ and $\mathscr{X}=\mathscr{D}$ we have a uniform tight rank-one IC-POVM. 
Embedded in $\Hd\cong\R^{d^2-1}$, the elements of a basis correspond to the vertices of a regular simplex in the 
$(d-1)$-dimensional subspace which they span. A complete set of MUBs corresponds to a maximal set of $d+1$ 
mutually orthogonal subspaces:
\begin{equation}
\frac{d}{d-1}\BraKetb{\pi(e_j^l)-I/d}{\pi(e_k^m)-I/d}\;=\;\left\{\begin{array}{ll} \frac{d\delta(j,k)-1}{d-1}\,, &\; l=m \\  0\,, &\; l\neq m \end{array}\right.\;. 
\end{equation}
Such IC-POVMs allow state determination via orthogonal measurements. The reconstruction formula is 
given by Eq.~(\ref{reconstructtight1}). Although constructions are known for prime-power dimensions~\cite{ivanovic,wootters} 
(see also \cite{alltop,calderbank,bandyopadhyay}), a complete set of MUBs is unlikely to exist in all dimensions.

Finally, let us rewrite the frame bound (Theorem~\ref{framethm}) for the context of quantum measurements. 

\begin{cor}\label{tightcor}
Let $F:\mathfrak{B}(\mathscr{X})\rightarrow\End(\C^d)$ be a POVM. Then 
\begin{equation}\label{framebound2}
\iint_\mathscr{X}\d\tau(x)\d\tau(y)\,\BraKetb{\Pt(x)}{\Pt(y)}^2 \;\geq\; 1+\frac{\big(\Tr(\mathcal{F})-1\big)^2}{d^2-1}\;,
\end{equation}
with equality iff $F$ is a tight IC-POVM.
\end{cor}
\begin{proof}
The frame bound [Eq.~(\ref{framebound})] takes the general form $\Tr(\mathcal{A}^2)\geq\big(\Tr(\mathcal{A})\big)^2/D$ where 
$D$ is the dimension of the operator space. Setting $\mathcal{A}=\mathcal{F}-\mathcal{I}/d$ and $D=d^2-1$ for $\Hd$ then 
gives Eq.~(\ref{framebound2}) [using Eq.~(\ref{identityeigenvector})].
\end{proof}
A POVM is a rank-one POVM if and only if $\Tr(\mathcal{F})=d$. With this value in the RHS of Eq.~(\ref{framebound2}) we 
recover the Welch bound (Theorem~\ref{welchthm}) for $t=2$.
\begin{cor}\label{tightcor2}
Let $F:\mathfrak{B}(\mathscr{X})\rightarrow\End(\C^d)$, $\mathscr{X}\subseteq\C P^{d-1}$, $\Pt\equiv\pi$, be a rank-one POVM. Then 
\begin{equation}\label{welchbound3}
\iint_\mathscr{X}\d\tau(x)\d\tau(y)\,\BraKetb{\pi(x)}{\pi(y)}^2 \;\geq\;\frac{2d}{d+1}\;,
\end{equation}
with equality iff $F$ is a tight IC-POVM.
\end{cor}

These corollaries tell us that tight IC-POVMs are those which minimize the average pairwise correlation in the POVD. 
An operational interpretation of this fact will be given in Sec.~\ref{optimalsec}. It is interesting to note that the 
above two examples of uniform tight rank-one IC-POVMs, SIC-POVMs and complete sets of MUBs, also minimize the maximal 
pairwise correlation. As spherical codes~\cite{conway} on the sphere of radius $\sqrt{(d-1)/d}$ in $\Hd$, 
SIC-POVMs saturate the simplex bound whilst complete sets of MUBs saturate the orthoplex bound (see e.g.~\cite{scott}).

\section{Optimal linear quantum state tomography}
\label{estimatesec}

Informationally complete quantum measurements are precisely those measurements which can be used for quantum state 
tomography. In this section we will show that, amongst all IC-POVMs, the tight rank-one IC-POVMs are the most robust 
against statistical error in the quantum tomographic process. We will also find that, for an arbitrary IC-POVM, 
the canonical dual frame with respect to the trace measure is the optimal dual frame for state reconstruction, 
thus confirming the approach of Sec.~\ref{icpovmsec}. These results, however, are shown only for the special case of 
{\em linear\/} quantum state tomography, which will be described later in this section.

Consider a state-reconstruction formula of the form 
\begin{equation}
\Ket{\rho} \;=\; \int_{\mathscr{X}} \d p(x)\,\Ketb{Q(x)} \;=\; \int_{\mathscr{X}} \BraKetb{\d F(x)}{\rho}\, \Ketb{Q(x)} \;,
\label{linearreconstruct}\end{equation}
where $Q:\mathscr{X}\rightarrow\End(\C^d)$ is an OVD. 
If this formula is to remain valid for all $\rho$, then we must have
\begin{equation}
\int_{\mathscr{X}} \Ketb{Q(x)}\Brab{\d F(x)} \;=\; \Is\;,
\label{reconstructidentityp}\end{equation}
which without loss of generality, can be rewritten as 
\begin{equation}
\int_{\mathscr{X}}\d\tau(x)\, \Ketb{Q(x)}\Brab{P(x)} \;=\; \Is\;,
\label{reconstructidentity}\end{equation}
where the POVD $P$ and trace measure $\tau$ are defined in Sec.~\ref{icpovmsec} [Eq.~(\ref{POVDeq})]. 
Equation~(\ref{reconstructidentity}) restricts $\{Q(x)\}_{x\in\mathscr{X}}$ to a dual frame of $\{P(x)\}_{x\in\mathscr{X}}$ with respect to the 
trace measure. Our first goal is to find the optimal dual frame.

It will be instructive to start with the special case of a discrete IC-POVM.
Suppose that we take $N$ random samples, $y_1,\dots,y_N$, from a countable set $\mathscr{X}$, where the
outcome $x$ occurs with some unknown probability $p(x)$. Our estimate for this probability is 
\begin{equation}
\hat{p}(x)\;=\;\hat{p}(x;y_1,\dots,y_N)\;\equiv\;\frac{1}{N}\sum_{k=1}^N\delta(x,y_k)\;, 
\label{pestimator}\end{equation}
which of course obeys the expectation $\mathrm{E}[\hat{p}(x)]=p(x)$. An elementary calculation shows that 
the expected covariance for $N$ samples is
\begin{equation}
\mathrm{E}\big[\big(p(x)-\hat{p}(x)\big)\big(p(y)-\hat{p}(y)\big)\big]\;=\;\frac{1}{N}\Big(p(x)\delta(x,y)-p(x)p(y)\Big)\;.
\label{covariance}\end{equation}

Now suppose that the $p(x)$ are outcome probabilities for an informationally complete quantum measurement of the state 
$\rho\in\Qd$. That is, $p(x)=\tr[F(x)\rho]$ where $\{F(x)\}_{x\in\mathscr{X}}\subset\End(\C^d)$ is a discrete IC-POVM. 
The error in our estimate of $\rho$, 
\begin{equation}
\hat{\rho} \;=\; \hat{\rho}(y_1,\dots,y_N) \;\equiv\; \sum_{x\in\mathscr{X}} \hat{p}(x;y_1,\dots,y_N)Q(x)\;,
\label{linearreconstructd}\end{equation}
as measured by the squared Hilbert-Schmidt (or Frobenius) distance, is 
\begin{equation}
\|\rho-\hat{\rho}\|^2 \;=\; \BraKet{\rho-\hat{\rho}}{\rho-\hat{\rho}}\;=\; \sum_{x,y\in\mathscr{X}}\big(p(x)-\hat{p}(x)\big)\big(p(y)-\hat{p}(y)\big)\BraKetb{Q(x)}{Q(y)}\;, 
\label{miscesteq}\end{equation}
which has the expectation
\begin{eqnarray}
\mathrm{E}\big[\|\rho-\hat{\rho}\|^2\big] &=& \frac{1}{N}\sum_{x,y\in\mathscr{X}}\big(p(x)\delta(x,y)-p(x)p(y)\big)\BraKetb{Q(x)}{Q(y)} \\ 
&=& \frac{1}{N}\bigg(\sum_{x\in\mathscr{X}}p(x)\BraKetb{Q(x)}{Q(x)}\,-\,\tr(\rho^2)\bigg) \\
&\equiv & \frac{1}{N}\Big(\Delta_p(Q)-\tr(\rho^2)\Big) \;, \label{esterror}
\end{eqnarray}
using Eq.~(\ref{covariance}) and then (\ref{linearreconstruct}). This expression is also a fitting description 
of the error from an IC-POVM with a continuum of measurement outcomes if we define 
\begin{equation}
\Delta_p(Q) \;\equiv\; \int_{\mathscr{X}} \d p(x)\,\BraKetb{Q(x)}{Q(x)} 
\label{Delta}\end{equation}
in general. This follows from the fact that a countable partition of the outcome set $\mathscr{X}$ allows any
continuous IC-POVM to be approximated by a discrete IC-POVM. Our estimate $\hat{p}$ 
remains a good approximation for the probability measure $p$, except now with $x$ and $y_1,\dots,y_N$ in 
Eq.~(\ref{pestimator}) indicating members of the partition. In the limit of finer approximating partitions we again arrive 
at Eq.~(\ref{esterror}) for the average error, but now with Eq.~(\ref{Delta}) for $\Delta_p(Q)$. 

Since we have no control over the purity of $\rho$, it is the quantity $\Delta_p(Q)$ in Eq.~(\ref{esterror}) which is 
now of interest. The IC-POVM which minimizes $\Delta_p(Q)$, and hence the error, will in general depend on the quantum state under 
examination. We thus set $\rho=\rho(\sigma,U)\equiv U\sigma U^\dag$, and now remove this dependence by taking the 
(Haar) average over all $U\in\mathrm{U}(d)$:
\begin{eqnarray}
\int_{\mathrm{U}(d)}\d\muu(U)\,\Delta_{p}(Q) &=& \int_{\mathrm{U}(d)}\d\muu(U)\,\int_{\mathscr{X}}\tr[\d F(x)U\sigma U^\dag]\BraKetb{Q(x)}{Q(x)} \\
&=& \frac{1}{d}\int_{\mathscr{X}}\tr[\d F(x)]\tr(\sigma)\BraKetb{Q(x)}{Q(x)} \\
&=& \frac{1}{d}\int_{\mathscr{X}}\d\tau(x)\,\BraKetb{Q(x)}{Q(x)} \\
&\equiv& \frac{1}{d}\,\Delta_\tau(Q)\;,
\end{eqnarray}
using Shur's Lemma for the integral and then setting $\tau\equiv\tr(F)$. The quantity $\Delta_\tau(Q)/d$ is 
the average value of $\Delta_p(Q)$ when $\rho$ is chosen randomly from an isotropic distribution 
in Euclidean space [via Eq.~(\ref{embedding})].

We will now minimize $\Delta_\tau(Q)$ over all choices for $Q$, while keeping the IC-POVM $F$ fixed. Our only constraint 
is that $\{Q(x)\}_{x\in\mathscr{X}}$ remains a dual frame to $\{P(x)\}_{x\in\mathscr{X}}$, so that the  
reconstruction formula [Eq.~(\ref{linearreconstruct})] remains valid for all $\rho$. The following lemma shows that the 
reconstruction OVD defined in Sec.~\ref{icpovmsec}, $\{R(x)\}_{x\in\mathscr{X}}$ [Eq.~(\ref{reconstructOVD})], is the 
optimal choice for the dual frame.

\begin{lem}\label{lemest1}
Let $\{A(x)\}_{x\in\mathscr{X}}\subseteq\End(\C^d)$ be an operator frame with respect to the measure $\alpha$. 
Then for all dual frames $\{B(x)\}_{x\in\mathscr{X}}$,
\begin{equation}
\Delta_\alpha(B) \;\equiv\; \int_{\mathscr{X}}\d\alpha(x)\,\BraKetb{B(x)}{B(x)} \;\geq\; \int_{\mathscr{X}}\d\alpha(x)\,\BraKetb{\tilde{A}(x)}{\tilde{A}(x)} \;\equiv\; \Delta_\alpha(\tilde{A}) \;,
\end{equation}
with equality only if $B\equiv\tilde{A}$, $\alpha$-almost everywhere, where $\{\tilde{A}(x)\}_{x\in\mathscr{X}}$ is the canonical dual frame.  
\end{lem}
\begin{proof}
Define $D\equiv B-\tilde{A}$ which satisfies 
\begin{eqnarray}
\int_{\mathscr{X}}\d\alpha(x)\,\Ketb{\tilde{A}(x)}\Brab{D(x)} &=& \int_{\mathscr{X}}\d\alpha(x)\,\Ketb{\tilde{A}(x)}\Brab{B(x)}\,-\,\int_{\mathscr{X}}\d\alpha(x)\,\KetBrab{\tilde{A}(x)} \\
&=& \int_{\mathscr{X}}\d\alpha(x)\,\mathcal{A}^{-1}\Ketb{A(x)}\Brab{B(x)}\,-\,\int_{\mathscr{X}}\d\alpha(x)\,\mathcal{A}^{-1}\KetBrab{A(x)}\mathcal{A}^{-1} \\
&=& \mathcal{A}^{-1}\Is\,-\,\mathcal{A}^{-1}\mathcal{A}\mathcal{A}^{-1} \\ 
&=& 0 \;, 
\end{eqnarray}
when $\{B(x)\}_{x\in\mathscr{X}}$ is a dual frame to $\{A(x)\}_{x\in\mathscr{X}}$ and $\{\tilde{A}(x)\}_{x\in\mathscr{X}}$ 
is the canonical dual frame, using Eq.'s~(\ref{dualframe}), (\ref{frameresolution}) and (\ref{framesuper}). Thus
\begin{equation}
\int_{\mathscr{X}}\d\alpha(x)\,\BraKetb{D(x)}{\tilde{A}(x)} \;=\; 0 \;,
\end{equation}
and
\begin{eqnarray}
\int_{\mathscr{X}}\d\alpha(x)\,\BraKetb{B(x)}{B(x)} &=& \int_{\mathscr{X}}\d\alpha(x)\,\BraKetb{\tilde{A}(x)}{\tilde{A}(x)}\,+\,\int_{\mathscr{X}}\d\alpha(x)\,\BraKetb{\tilde{A}(x)}{D(x)} \\ 
&& \,+\,\int_{\mathscr{X}}\d\alpha(x)\,\BraKetb{D(x)}{\tilde{A}(x)}\,+\,\int_{\mathscr{X}}\d\alpha(x)\,\BraKetb{D(x)}{D(x)} \\
&=& \int_{\mathscr{X}}\d\alpha(x)\,\BraKetb{\tilde{A}(x)}{\tilde{A}(x)}\,+\,\int_{\mathscr{X}}\d\alpha(x)\,\BraKetb{D(x)}{D(x)} \\
&\geq & \int_{\mathscr{X}}\d\alpha(x)\,\BraKetb{\tilde{A}(x)}{\tilde{A}(x)} \;,
\end{eqnarray}
with equality if and only if $D\equiv 0$, $\alpha$-almost everywhere.
\end{proof}

Setting $A\equiv P$ and $\alpha\equiv\tau$ in Lemma~\ref{lemest1} confirms the reconstruction method presented in 
Sec.~\ref{icpovmsec} [Eq.'s~(\ref{framesuperoperator}), (\ref{reconstructOVD}) and (\ref{reconstruct})]. Notice that 
we can retain the dependence on $\rho$ by simply replacing $\tau$ by $pd$ in these formulae. An adaptive reconstruction 
method might make use of this fact. Equation~(\ref{framesuperoperatorinverse}) shows that 
$\Delta_\tau(R)=\Tr(\mathcal{F}^{-1})$. This quantity will now be minimized over all IC-POVMs. 

\begin{lem}\label{lemest2}
Let $F:\mathfrak{B}(\mathscr{X})\rightarrow\End(\C^d)$ be an IC-POVM. Then
\begin{equation}
\Tr(\mathcal{F}^{-1}) \;\geq\; d\big(d(d+1)-1\big)\;,
\end{equation}
with equality iff $F$ is a tight rank-one IC-POVM.
\end{lem}
\begin{proof}
We will minimize the quantity
\begin{equation}
\Tr(\mathcal{F}^{-1}) \;=\; \sum_{k=1}^{d^2} \frac{1}{\lambda_k} \;,
\label{lemest2proof1}\end{equation}
where $\lambda_1,\dots,\lambda_{d^2}>0$ denote the left-right eigenvalues of $\mathcal{F}$. These eigenvalues satisfy the constraint
\begin{equation}
\sum_{k=1}^{d^2} \lambda_k \;=\; \Tr(\mathcal{F}) \;=\; \int_{\mathscr{X}} \d\tau(x)\,\BraKetb{P(x)}{P(x)} \;\leq\; \int_{\mathscr{X}} \d\tau(x) \;=\; d \;,
\end{equation} 
since $\tr(P^2)\leq 1$, $\tau$-almost everywhere. We know, however, that the identity operator is always a 
left-right eigenvector of $\mathcal{F}$ with unit eigenvalue [Eq.~(\ref{identityeigenvector})]. Thus we in fact have $\lambda_1=1$ say, and then 
$\sum_{k=2}^{d^2} \lambda_k\leq d-1$. Under this latter constraint it is straightforward to show that the RHS of 
Eq.~(\ref{lemest2proof1}) takes its minimum value if and only if $\lambda_2=\dots=\lambda_{d^2}=(d-1)/(d^2-1)=1/(d+1)$, 
or equivalently,
\begin{equation}
\mathcal{F} \;=\; 1\cdot\frac{\mathcal{I}}{d} \,+\, \frac{1}{d+1}\cdot\Ptr \;=\; \frac{\Is + \mathcal{I}}{d+1} \;,
\end{equation}
since the subspace of traceless operators is $\mathcal{F}$-invariant [Eq.~(\ref{POVMdecomp})]. Therefore, 
by Proposition~\ref{tight1prp}, $\Tr(\mathcal{F}^{-1})$ takes its minimum value if and only if $F$ is a tight 
rank-one IC-POVM. The minimum is $\Tr(\mathcal{F}^{-1}) = 1\cdot 1+(d+1)\cdot (d^2-1)=d\big(d(d+1)-1\big)$.
\end{proof}

We have thus confirmed that it is optimal to use a tight rank-one IC-POVM for quantum state tomography. The optimal 
state-reconstruction formula is then given by Eq.~(\ref{reconstructtight1}). 
Before stating these results in a theorem, let us first fully clarify the assumptions that have allowed us to draw this conclusion. 
First of all, we have chosen the Hilbert-Schmidt metric to measure distances in $\Qd$ [see Eq.~(\ref{miscesteq})]. There 
are other choices to consider and some of these are no doubt more appropriate in the context of quantum states. For 
example, we could instead quantify the error in $\hat{\rho}$ with the Bures metric \cite{bures,uhlmann} 
$d_\textrm{B}(\rho,\hat{\rho})^2\equiv 2-2\tr\sqrt{\sqrt{\rho}\hat{\rho}\sqrt{\rho}}$, or, although not strictly a metric, 
the relative entropy $S(\rho||\hat{\rho})\equiv \tr(\rho\log\rho-\rho\log\hat{\rho})$. 
These choices, however, proved too cumbersome to warrant a detailed investigation in the current article.

We have also made assumptions about the procedure for state reconstruction. This can be explained as follows.
For an informationally complete POVM, $F$ say, every quantum state is uniquely identified by its 
measurement statistics. This does not mean, however, that all points on the probability simplex, 
$\int_\mathscr{X}\d p(x)=1$, describe valid outcome statistics for a measurement (with IC-POVM $F$) of a 
quantum state. Due to the possible overcompleteness of a POVM, there can be many choices for the estimate statistics 
$\hat{p}$ (just as there were many choices for the reconstruction OVD $Q$) which satisfy 
$\int_\mathscr{X} \d \hat{p}(x)Q(x)=\rho$ [Eq.~(\ref{linearreconstruct})] for some fixed $\rho\in\Qd$. The state's actual 
measurement statistics, $p\equiv\tr(F\rho)$, are only but one 
of these choices. Additionally, for some choices of the estimate statistics we might not even have 
$\int_\mathscr{X} \d \hat{p}(x)Q(x)\in\Qd$. 

Suppose, for example, that we have a finite IC-POVM with $|\mathscr{X}|=d^2+K$ possible measurement outcomes. We know that every POVM satisfies 
the normalization constraint, $\sum_{x\in\mathscr{X}} F(x)=I$, which implies normalization of the statistics: 
$\sum_{x\in\mathscr{X}}p(x)=1$. Our previous estimate [Eq.~(\ref{pestimator})] satisfies this constraint. It does 
not, however, incorporate any additional constraints specific to the particular choice of IC-POVM.
Embedding the POVM elements in $\Hd$ shows that there will be a further $K$ linear constraints of the form
\begin{equation}
\sum_{x\in\mathscr{X}} c_k(x) F(x) \; = \; 0\;, \quad \text{which imply that} \quad
\sum_{x\in\mathscr{X}} c_k(x) p(x) \; = \; 0 \qquad \big(c_k(x)\in\R\,,\;k=1,\dots, K\big)\;.
\label{Kconstraints}\end{equation}
The intersection of the probability simplex in $\R^{d^2+K}$ with the subspace perpendicular to the $K$ vectors 
$\{c_k(x)\}_{x\in\mathscr{X}}$ forms the subset of statistics which are isomorphic, under the mapping  
$p=\tr(FA)\rightarrow A$, to the normalized Hermitian operators in $\End(\C^d)$. We can thus excise all unphysical 
estimate statistics which duplicate valid measurement statistics by taking these extra constraints into account. After $N$ 
measurements, with results $y_1,\dots,y_N$, the most appropriate choice for $\hat{p}$ will be the maximum-likelihood 
estimate under these constraints, i.e. that which maximizes $\textrm{Prob}(p)=\prod_{x\in\mathscr{X}} p(x)^{n(x)}$, 
where $n(x)\equiv\sum_{k=1}^N\delta(x,y_k)$. Under the normalization constraint only, it is straightforward to recover 
$\hat{p}(x)=n(x)/N$ [Eq.~(\ref{pestimator})]; under both the normalization and additional constraints, however, this 
nonlinear optimization problem becomes difficult to solve analytically. One exception is an IC-POVM consisting of $d+1$ MUBs 
[Eq.~(\ref{MUBs})], in which case the $K=d$ additional constraints [$c_k(e_j^l)=(d+1)\delta(k,l)-1$ in Eq.~(\ref{Kconstraints})] single out 
$\hat{p}(e_j^l) =n(e_j^l)/\big[(d+1)\sum_{k=1}^d n(e_k^l)\big]$ for the maximum-likelihood estimate, as one should expect. 
This means we should treat the outcome probabilities as if they came from $d+1$ separate orthogonal measurements, each 
corresponding to one of the bases. In the special case of a {\em minimal\/} IC-POVM (i.e $|\mathscr{X}|=d^2$) there are no additional constraints and
Eq.~(\ref{pestimator}) is the best estimate for the outcome statistics. For this reason minimal IC-POVMs should 
be preferred over other IC-POVMs. Lemma~\ref{lemest1} is then redundant since the canonical dual frame is 
the unique dual frame, namely the dual basis. In general, only when all $K+1$ linear constraints 
are taken into account is Lemma~\ref{lemest1} unnecessary and the particular choice of reconstruction formula 
unimportant. 

By taking the maximum-likelihood estimate under the normalization and all $|\mathscr{X}|-d^2$ additional linear 
constraints we can remove the redundancy in the estimate statistics. There may still remain unphysical  
statistics however. If $\rho$ is pure, or if $N$ is not large enough, then $\int_\mathscr{X} \d \hat{p}(x)Q(x)$
may not be a positive operator under the linear constraints, and thus, not a quantum state. To overcome this problem we 
must instead apply the single nonlinear constraint that $\hat{p}\in\{\tr(F\rho)|\rho\in\Qd\}$, and again take the 
maximum-likelihood estimate. 

To show that the tight rank-one IC-POVMs are optimal for quantum state tomography we have ignored all additional 
linear and nonlinear constraints on the estimate statistics, and simply taken Eq.~(\ref{pestimator}) for $\hat{p}$, 
with a reconstruction formula in the form of Eq.~(\ref{linearreconstruct}). Although this simplification will likely 
lead to less than optimal estimates of the quantum state, the inclusion of all possible constraints 
on $\hat{p}$ for the maximum-likelihood estimation, or only the linear constraints, makes any generalization of our results considerably more difficult.
In this article we will thus only claim that tight rank-one IC-POVMs are optimal for {\em linear\/} quantum state tomography, with 
the term `linear' referring to the previous simplified state-reconstruction procedure, i.e. without 
the nonlinear optimization needed for maximum-likelihood estimation under the additional constraints. This result 
is summarized in the following theorem.

\begin{thm}\label{thmest}
Let $F:\mathfrak{B}(\mathscr{X})\rightarrow\End(\C^d)$ be an IC-POVM and let $\rho=\rho(\sigma,U)\equiv U\sigma U^\dag$ 
for some fixed quantum state $\sigma\in\Qd$. Then
\begin{equation}
e_\mathrm{av}^{(F,Q)}(\sigma) \;\equiv\; \int_{\mathrm{U}(d)}\d\muu(U)\:\mathrm{E}\big[\|\rho-\hat{\rho}\|^2\big] \;\geq\; \frac{1}{N}\bigg(\frac{1}{d}\Tr(\mathcal{F}^{-1})-\tr(\sigma^2)\bigg) \;\geq\; \frac{1}{N}\Big(d(d+1)-1-\tr(\sigma^2)\Big)
\label{thmesteq}\end{equation}
for all reconstruction OVDs $Q:\mathscr{X}\rightarrow\End(\C^d)$ which are dual frames to $P$, 
where $\hat{\rho}=\hat{\rho}(\sigma,U;y_1,\dots,y_N)$ is a linear tomographic estimate of $\rho$ given $N$ 
measurement outcomes $y_1,\dots,y_N$ [Eq.'s~(\ref{pestimator}) and (\ref{linearreconstructd})] and the expectation is over 
these outcomes. Furthermore, equality in the LHS of Eq.~(\ref{thmesteq}) occurs iff $Q\equiv R$, $\tau$-almost everywhere, and equality 
in the RHS of Eq.~(\ref{thmesteq}) occurs iff $F$ is a tight rank-one IC-POVM. 
\end{thm}

We can also consider the worst-case expectation in the error. The average then provides a lower bound:
\begin{equation}
e_\mathrm{wc}(\sigma) \;\equiv\; \sup_{U\in\mathrm{U}(d)}\:\mathrm{E}\big[\|\rho-\hat{\rho}\|^2\big] \;\geq\; e_\mathrm{av}(\sigma) \;\geq\; \frac{1}{N}\Big(d(d+1)-1-\tr(\sigma^2)\Big)\;.
\label{estwcineq}\end{equation}
Notice, however, that if $R=(d+1)P-I=(d+1)\pi-I$, as defined for a tight rank-one IC-POVM 
[Eq.'s~(\ref{tightreconstructionOVD}) and (\ref{tightrank1ICPOVMa}) with $P=\pi$], then 
$\BraKet{R}{R}=\tr(R^2)=d(d+1)-1$, $\tau$-almost everywhere. Consequently, returning to Eq.~(\ref{esterror}) but 
now with $Q=R$ and $\rho=\rho(\sigma,U)\equiv U\sigma U^\dag$, we see that regardless of the choice of 
$U\in\mathrm{U}(d)$, for a tight rank-one IC-POVM we always have
\begin{eqnarray}
e(\sigma,U) \;\equiv\;\mathrm{E}\big[\|\rho-\hat{\rho}\|^2\big] &=& \frac{1}{N}\bigg(\int_{\mathscr{X}}\d p(x)\,\BraKetb{R(x)}{R(x)}\,-\,\tr(\sigma^2)\bigg) \\
&=& \frac{1}{N}\bigg(\big(d(d+1)-1\big)\int_{\mathscr{X}}\d p(x)\,-\,\tr(\sigma^2)\bigg) \\
&=& \frac{1}{N}\Big(d(d+1)-1-\tr(\sigma^2)\Big)
\end{eqnarray}
when Eq.~(\ref{reconstructtight1}) is used for state reconstruction. The above inequality [Eq.~(\ref{estwcineq})] and 
this last fact implies the following corollary to Theorem~$\ref{thmest}$.

\begin{cor}\label{corest}
Let $F:\mathfrak{B}(\mathscr{X})\rightarrow\End(\C^d)$ be an IC-POVM and let $\rho=\rho(\sigma,U)\equiv U\sigma U^\dag$ 
for some fixed quantum state $\sigma\in\Qd$. Then
\begin{equation}
e_\mathrm{wc}^{(F,Q)}(\sigma) \;\equiv\; \sup_{U\in\mathrm{U}(d)}\:\mathrm{E}\big[\|\rho-\hat{\rho}\|^2\big] \;\geq\; \frac{1}{N}\Big(d(d+1)-1-\tr(\sigma^2)\Big)
\label{coresteq}\end{equation}
for all reconstruction OVDs $Q:\mathscr{X}\rightarrow\End(\C^d)$ which are dual frames to $P$, where $\hat{\rho}=\hat{\rho}(\sigma,U;y_1,\dots,y_N)$ is a 
linear tomographic estimate of $\rho$ given $N$ measurement outcomes $y_1,\dots,y_N$ [Eq.'s~(\ref{pestimator}) and (\ref{linearreconstructd})] and the expectation is over 
these outcomes. Furthermore, equality in Eq.~(\ref{coresteq}) occurs iff $Q\equiv R$, $\tau$-almost everywhere, and $F$ is a tight rank-one IC-POVM. 
\end{cor}

Tight rank-one IC-POVMs are thus optimal for linear quantum state tomography in both an average and worst-case sense.
In fact, they form the unique class of POVMs capable of achieving
\begin{equation}
e_\mathrm{wc}(\sigma) \;=\; e_\mathrm{av}(\sigma) \;=\; e(\sigma,U)  \;=\; \frac{1}{N}\Big(d(d+1)-1-\tr(\sigma^2)\Big)\;.
\end{equation}

The type of quantum state tomography considered in this section was based on nonadaptive sequential measurements 
on copies of the quantum state. This restriction is detrimental to the tomographic process. Given multiple copies of 
a state, there exist joint measurements on these copies which will outperform any of the measurements considered above 
(see e.g.~\cite{vidal}). In the next section, however, we will show that the tight rank-one IC-POVMs form the unique 
class of POVMs which are optimal for state estimation, if given only a single copy of a pure quantum state.

\section{Optimal measurement-based cloners}
\label{optimalsec}

A natural way of assessing the capability of a measuring instrument for state estimation is to consider 
it in the role of a cloning machine~\cite{gisin,massar,derka,latorre,bruss,hayashi}. A single copy of an unknown pure quantum state $\psi\in\C P^{d-1}$ 
is the input to this device, while the output is a finite number of approximate copies of $\psi$, or in the case of a measurement, 
an infinite supply of approximate copies described by a single mixed quantum state. This estimate  
will in general depend on the measurement result. For outcome $x$ we will denote the device's output state
by $\hat{\rho}(x)\in\Qd$. The probability of confirming $\hat{\rho}(x)$ to be $\pi(\psi)$ is then given by the fidelity, 
$f(\psi,x)\equiv\bra{\psi}\hat{\rho}(x)\ket{\psi}$. The average fidelity over all measurement outcomes,
\begin{equation}
f(\psi)\;\equiv\;\int_\mathscr{X} \tr\!\big[\d F(x)\pi(\psi)\big] f(\psi,x) \;=\;\int_\mathscr{X}\d\tau(x)\, \bra{\psi}\Pt(x)\ketbra{\psi}\hat{\rho}(x)\ket{\psi}\;,
\end{equation}
is the probability that the POVM $F$, together with the estimate state $\hat{\rho}$, successfully clones $\psi$. 
Maximized over all choices for $\hat{\rho}$, this quantity might be interpreted as an operational measure 
of knowledge (about $\psi$) gained from the measurement.  
For the purposes of this section we will call the pair $(F,\hat{\rho})$ a {\em measurement-based cloning strategy\/}.

Consider the average success probability for such strategies: 
\begin{eqnarray}
f_\textrm{av} &\equiv& \int_{\C P^{d-1}}\d\muu(\psi)\,f(\psi) \\ 
&=& \int_{\C P^{d-1}}\d\muu(\psi)\int_{\mathscr{X}}\d\tau(x)\, \tr\!\big[\pi(\psi)^{\otimes 2}\,\cdot\, \Pt(x)\otimes\hat{\rho}(x)\big] \\
&=& \frac{2}{d(d+1)}\int_\mathscr{X}\d\tau(x)\, \tr\!\big[\Pisym^{(2)}\,\cdot\, \Pt(x)\otimes\hat{\rho}(x)\big] \\
&=& \frac{1}{d(d+1)}\int_\mathscr{X}\d\tau(x)\,\big(1+\tr[\Pt(x)\hat{\rho}(x)]\big) \\
&\leq& \frac{2}{d+1} \;.
\end{eqnarray}
Here we have used Lemma~\ref{lemsym} and then the identity $2\tr\!\big(\Pisym^{(2)}\cdot A\otimes B\big)=\tr(A)\tr(B)+\tr(AB)$.
Equality will occur if and only if $\tr(\Pt\hat{\rho})=1$, $\tau$-almost everywhere, in which case we must have 
$\hat{\rho}=\Pt=\pi$, where we now consider $\mathscr{X}\subseteq\C P^{d-1}$. Thus $F$ is capable of achieving the 
maximum possible average success probability if and only if it is a rank-one POVM. It is no surprise that the best 
choice for the estimate state is then given by the POVD. 

But can we ask for more from the measuring instrument? Let us instead maximize the worst-case 
success probability. This quantity may be thought of as a guarantee on the success rate. The average success 
probability provides an upper bound:
\begin{equation}
f_\textrm{wc} \;\equiv\; \inf_{\psi\in\C P^{d-1}}f(\psi) \;\leq\; f_\textrm{av} \;\leq\; \frac{2}{d+1}\;.
\end{equation}
Now consider the conditions upon which equality is achieved. First of all we need $f_\textrm{av}=2/(d+1)$, 
and thus, we require a rank-one POVM $\Pt=\pi$ with the estimate state $\hat{\rho}=\pi$. If additionally we have 
$f_\textrm{wc}=f_\textrm{av}$ then the variance in the success probability must necessarily vanish, or equivalently, 
$\int_{\C P^{d-1}}\d\muu(\psi)\,f(\psi)^2=f_\textrm{av}^2=4/(d+1)^2$. The second moment may be 
calculated in a similar manner to the first:
\begin{eqnarray}
 \int_{\C P^{d-1}}\d\muu(\psi)\,f(\psi)^2 &=& \int_{\C P^{d-1}}\d\muu(\psi)\iint_{\mathscr{X}}\d\tau(x)\d\tau(y)\, \tr\big[\pi(\psi)^{\otimes 4}\,\cdot\, \pi(x)^{\otimes 2}\otimes\pi(y)^{\otimes 2}\big] \\
&=& \frac{24}{d(d+1)(d+2)(d+3)}\iint_\mathscr{X}\d\tau(x)\d\tau(y)\, \tr\big[\Pisym^{(4)}\,\cdot\, \pi(x)^{\otimes 2}\otimes\pi(y)^{\otimes 2}\big] \\
&=& \frac{4}{d(d+1)(d+2)(d+3)}\iint_\mathscr{X}\d\tau(x)\d\tau(y)\, \big(1+4\tr[\pi(x)\pi(y)]+\tr[\pi(x)\pi(y)]^2\big) \\
&=& \frac{4}{d(d+1)(d+2)(d+3)}\bigg(d^2+4d+\iint_\mathscr{X}\d\tau(x)\d\tau(y)\,|\braket{x}{y}|^4\bigg)\;.
\end{eqnarray}
Here we have again used Lemma~\ref{lemsym} and then a similar identity to the above, except this time with 
$4!$ terms. Given the second moment, one can easily check that the condition for zero variance is equivalent to
\begin{equation}
\iint_\mathscr{X}\d\tau(x)\d\tau(y)\,\BraKetb{\pi(x)}{\pi(y)}^2\;=\;\iint_\mathscr{X}\d\tau(x)\d\tau(y)\,|\braket{x}{y}|^4\;=\;\frac{2d}{d+1}\;,
\end{equation}
which by Corollary~\ref{tightcor2}, implies that $F$ is a tight rank-one IC-POVM. This condition is also sufficient. It is 
straightforward to confirm that for tight rank-one IC-POVMs, $f(\psi)=2/(d+1)$ independent of $\psi$.

We have just shown that the worst-case success probability, for a measuring instrument in the 
role of a cloning machine, can take its maximum value if and only if the corresponding POVM is a tight rank-one IC-POVM. 
In fact, the tight rank-one IC-POVMs form the unique class of POVMs capable of achieving $f_\textrm{wc}=f_\textrm{av}=f(\psi)=2/(d+1)$. 
It is in this sense that a tight rank-one IC-POVM can be claimed optimal for state estimation. 
Notice that, unlike a generic rank-one POVM, a strategy based on a tight rank-one IC-POVM outputs on average an 
isotropically unbiased estimate of the input state:  
\begin{eqnarray}
\mathrm{E}\big[\hat{\rho}(x)\big]\;=\;\int_\mathscr{X} \tr\!\big[\d F(x)\pi(\psi)\big] \hat{\rho}(x) & = & \int_\mathscr{X}\d\tau(x)\, \bra{\psi}\pi(x)\ket{\psi}\pi(x) \\
&=& \int_\mathscr{X}\d\tau(x)\, \Ketb{\pi(x)}\BraKetb{\pi(x)}{\pi(\psi)} \\
&=& \mathcal{F}\Ketb{\pi(\psi)} \\ 
&=& \frac{I+\ketbra{\psi}}{d+1}\;,
\label{unbiasedness}\end{eqnarray}
where we have used Proposition~\ref{tight1prp}. Only the tight rank-one IC-POVMs satisfy Eq.~(\ref{unbiasedness}),
which could be taken as a defining property. Let us now restate the above facts formally in a theorem.

\begin{thm}\label{clonethm}
Let $(F,\hat{\rho})$ be a measurement-based cloning strategy with POVM $F:\mathfrak{B}(\mathscr{X})\rightarrow\End(\C^d)$. 
Then  
\begin{equation}
f_\mathrm{wc}^{(F,\hat{\rho})} \;\equiv\; \inf_{\psi\in\C P^{d-1}}\; \int_\mathscr{X} \tr\!\big[\d F(x)\pi(\psi)\big]\tr\!\big[\hat{\rho}(x)\pi(\psi)\big] \;\leq\; \frac{2}{d+1} \;,
\end{equation}
with equality if and only if $F$ is a tight rank-one IC-POVM and $\hat{\rho}=P$.  
\end{thm}

This theorem is in fact a special case of the results of Hayashi {\it et al.}~\cite{hayashi} (see also~\cite{latorre}). If instead $N$ copies of $\psi$ are 
given, then the optimal joint measurement on these copies that maximizes the average success probability is defined by an 
$N$-design. The success probability then increases to $f_\textrm{av}=(N+1)/(N+d)$~\cite{bruss}. 
The measurement that maximizes the worst-case success probability is instead defined by an $(N+1)$-design, 
in which case we have $f_\textrm{wc}=f_\textrm{av}=f(\psi)=(N+1)/(N+d)$. The extension to mixed states for this type of quantum 
state estimation has been considered by Vidal {\it et al.}~\cite{vidal}.

\section{Conclusion}
\label{concludesec}

In this article we have introduced a special class of informationally complete POVMs which, in analogy to a similar concept in 
frame theory, are named {\em tight IC-POVMs\/}. Embedded as a tight frame in the vector space of all traceless Hermitian operators, 
which is the natural place to study a quantum state, a tight IC-POVM is as close as possible to an orthonormal basis. It is in this sense 
that the tight IC-POVMs can be promoted as being special amongst all IC-POVMs. They allow painless quantum state tomography through a 
particularly simple state-reconstruction formula [Eq.~(\ref{reconstructtight})]. The rank-one members of this class have the minimum 
average pairwise correlation in the POVD for any rank-one POVM (Corollary~\ref{tightcor2}) and thus form the family of optimal measurement-based cloners 
(Theorem~\ref{clonethm}). They are also the best choice for linear quantum state tomography (Theorem~\ref{thmest} and Corollary~\ref{corest}). 
The outstanding choice amongst all tight rank-one IC-POVMs are the unique minimal members, the SIC-POVMs~\cite{renes}. These POVMs are the 
only equiangular tight rank-one IC-POVMs, minimize the maximal pairwise correlation in the POVD, and can thus be considered the closest, 
now amongst all tight rank-one IC-POVMs, to an orthonormal basis.

\begin{acknowledgments} 
This work has been supported by CIAR, CSE, iCORE and MITACS. 
\end{acknowledgments}

\end{document}